# Research on Adverse Drug Reaction Prediction Model Combining Knowledge Graph Embedding and Deep Learning

Yufeng Li*
University of Southampton, Southampton, UK,
*liyufeng0913@gmail.com

Wenchao Zhao
University of Science and Technology of China,
Anhui, China, ywzhaohong@gmail.com

Bo Dang
Computer Science, San Francisco Bay University Fremont, CA, US
dangdaxia@gmail.com

Xu Yan
Trine University, Phoenix, Arizona, US
yancontiue@gmail.com

Min Gao
Trine University, Allen Park, Michigan, US
mingao4460@gmail.com

Weimin Wang
The Hong Kong University of Science and Technology, Hongkong, China
wangwaynemin@gmail.com

Mingxuan Xiao
SouthWest JiaoTong University, Chengdu, Sichuan, China
553556963albert@gmail.com

*Abstract*—In clinical treatment, identifying potential adverse reactions of drugs can help assist doctors in making medication decisions. In response to the problems in previous studies that features are high-dimensional and sparse, independent prediction models need to be constructed for each adverse reaction of drugs, and the prediction accuracy is low, this paper develops an adverse drug reaction prediction model based on knowledge graph embedding and deep learning, which can predict experimental results. Unified prediction of adverse drug reactions covered. Knowledge graph embedding technology can fuse the associated information between drugs and alleviate the shortcomings of high-dimensional sparsity in feature matrices, and the efficient training capabilities of deep learning can improve the prediction accuracy of the model. This article builds an adverse drug reaction knowledge graph based on drug feature data; by analyzing the embedding effect of the knowledge graph under different embedding strategies, the best embedding strategy is selected to obtain sample vectors; and then a convolutional neural network model is constructed to predict adverse reactions. The results show that under the DistMult embedding model and 400-dimensional embedding strategy, the convolutional neural network model has the best prediction effect; the average accuracy, $F_1$ score, recall rate and area under the curve of repeated experiments are better than the methods reported in the literature. The obtained prediction model has good prediction accuracy and stability, and can provide an effective reference for later safe medication guidance.

## I. INTRODUCTION

Adverse Drug Reaction (ADR) is an important global public health problem and one of the major causes of death[1]. The World Health Organization vigibase database records approximately 23 million ADRs, of which 43,685 are fatal ADRs. Common drugs that cause ADRs are anti-tumor drugs, neurological drugs and cardiac-related drugs[2]. In the United States, more than 2 million hospitalized patients suffer from severe ADR every year, causing economic losses of US$528.4 billion, accounting for about 16% of the total medical expenditures that year[3]. Although drugs undergo rigorous testing before being approved for marketing, many serious ADRs do not appear until some time after the drug is on the market due to limited sample numbers and trial time constraints. Additionally, some ADR-related hospitalizations may be avoided by avoiding inappropriate prescribing[4]. Therefore, how to effectively identify and predict potential adverse reactions of drugs, prevent the occurrence of ADRs, reduce economic losses, and improve the rationality and safety of clinical medication has always been a research focus in the current field of smart health care. Deep learning can be used in multiple intelligent fields, such as image change detection[27] and recognition[28], and even AI text detection[26]. This project is used in the prediction field. Based on this, this project develops an ADR prediction model based on knowledge graph embedding and deep learning, and conducts comparative analysis with a variety of commonly used benchmark models and existing research results, while testing the effectiveness and stability of this prediction model.

## II. RELATED RESEARCH WORK

According to the definition of the World Health Organization, adverse drug reactions refer to harmful and unexpected reactions that occur during the use of normal doses of drugs for the prevention, diagnosis, treatment of diseases or the regulation of physiological functions and are unrelated to the purpose of medication[5]; and ADRs may It is the result of the reaction between pharmaceutical chemicals and proteins[6].

Early research on ADRs was mainly based on clinical case data from Spontaneous Reporting Systems (SRSs), using methods such as proportional imbalance analysis to evaluate the correlation and causality between drugs and ADRs to mine relevant adverse drug reaction signals. However, the data of SRSs are often incomplete or inaccurate, which may lead to biased research results. In addition, the limited amount of data and the lack of in-depth data mining make early research conclusions based on simple statistical methods unconvincing[7]. As artificial intelligence technology continues to develop and the amount of medical data continues to grow, on the one hand, researchers are mining potential adverse reactions of drugs based on text data such as literature and ADR reports, combined with natural language processing technology[8]; on the other hand, based on the chemistry and biology of drugs, and phenotypic characteristics, and use machine learning or deep learning methods to conduct ADR prediction research[9]. Research based on text mining is often used to identify and monitor relevant ADRs, which assumes that relevant ADRs have appeared, but cannot predict potential ADRs of drugs; while ADR prediction research based on drug features and machine learning helps to explore unknown ADRs of drugs, which it is also the research topic of this project. Machine learning related methods can improve the ADR prediction effect, but there are still key points that can be improved in these studies: (1) the correlation between drugs is not considered, which may lead to the loss of useful information; (2) the use of a large amount of one-hot encoded feature data, However, it is difficult to reduce the dimensionality of high-dimensional sparse feature matrices, and the model calculation efficiency is low; (3) Most of them need to build separate classifiers for each ADR. Knowledge Graph (KG), a special network structure composed of nodes and relationships and its embedding technology, captures the unstructured semantic relationships between feature entities by embedding entities into a continuous low-dimensional feature space. Fusion and calculation between information can effectively alleviate the computational inefficiency problem caused by high-dimensional sparse feature data and improve the prediction performance of the classifier[10].

In recent years, knowledge graphs and their embedding technology have been gradually used in knowledge discovery and knowledge base construction in the field of drug research. These studies obtain drug feature data, build knowledge graphs containing different types of nodes, and combine knowledge graph embedding technology with classification models. Target predictions for related research topics. Based on KG's ADR prediction, related typical research is shown in Table 1. Through literature review, the current research still has the following key points that need to be improved: (1) Use "Drug-ADR" combinations that do not appear in KG as negative samples for the ADR prediction model, but "Drug-ADR" combinations that do not exist in KG may It's just that it has not been discovered yet; (2) a simple machine learning model is used; (3) the number of drugs covered is small, the characteristics are limited to drug targets and indications, and important information such as enzymes and carrier proteins have not been previously reported used in research.

Table 1. Relevant Typical Studies

| Research | Year | Number of drugs | Feature Category | Data source | Classification model |
|---|---|---|---|---|---|
| Joshi et al[11] | 2022 | 7219 | ADR, Indication, Target, Pathway, Gene | DrugBank, SIDER | DNN |
| Zhang et al[12] | 2021 | 3632 | Target, Indication, ADR | DrugBank, SIDER | LR |
| Wang et al[13] | 2021 | 1806 | Tumor, Biomarker, ADR | MEDLINE | NB |
| Dey et al[14] | 2018 | 1430 | Chemical structure, Side effect | SIDER, PubChem | CNN |
| Bean et al[15] | 2017 | 524 | Target, Indication, ADR | DrugBank, SIDER | LR |

Based on this, this article uses a method that combines knowledge graph embedding and deep learning to achieve ADR prediction. In addition to targets and indications, it also integrates enzyme and carrier protein information to construct a knowledge graph; and develops a powerful deep neural network to improve ADR prediction performance.

### III. DATA AND METHODOLOGY

In the method proposed in this paper, referring to existing literature[11][12], the side effects of drugs are regarded as ADRs. In view of the fact that combining the biological characteristics and phenotypic characteristics of drugs can improve the performance of the ADR prediction model, biological characteristics and indications such as target, transporter, and enzyme were selected from the DrugBank (v5.18)[16] and SIDER (v4.1)[17] databases respectively, and adverse reactions (ADR) and other phenotypic characteristics, and drugs (drug) as knowledge graph entity nodes. Then, in order to avoid the heavy workload of building a separate classifier for each ADR, ADR prediction is treated as a unified two-classification problem, and the "Drug-ADR" combination and the "Drug-Indication" combination are used as the positive components of the classification model respectively. For samples and negative samples, the sample labels are recorded as "1" and "0" respectively. From this, an ADR prediction model based on knowledge graph embedding and deep learning was developed, and the stability of the Convolutional Neural Networks (CNN) model was tested through five repeated experiments. Finally, drug-induced renal function injury was used as an example to predict, and the validity of the model prediction was verified through real-world data. The specific research ideas are shown in Figure 1.

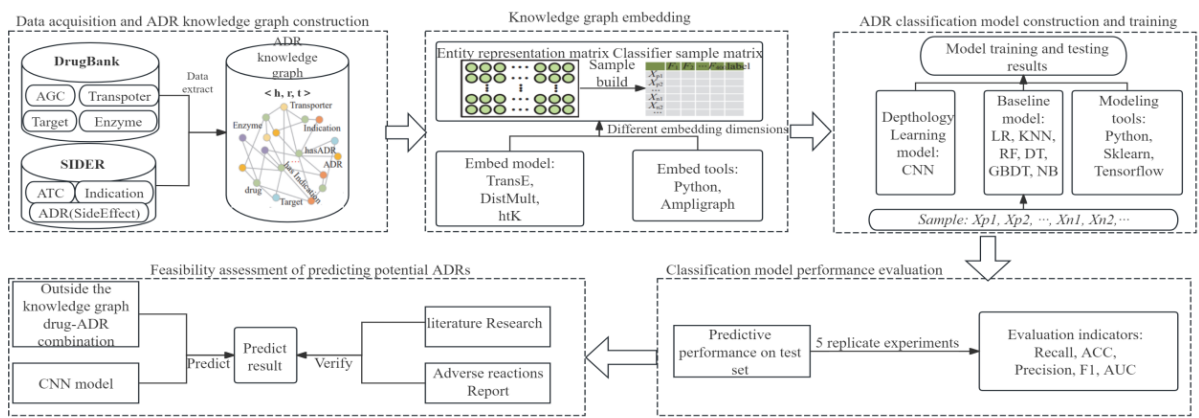

Figure 1. Framework of ADR prediction

### A. Data Sources and Knowledge Graph Construction

The DrugBank database covers a wealth of biological and chemical informatics resources, and the SIDER database includes 1,500 drugs and more than 6,100 side effects. By downloading the xml data file in DrugBank and the tsv file in SIDER, use the Python program to parse and obtain the relevant characteristic data of the drug. Integrate relevant data from two databases according to the Anatomical Therapeutic Chemical (ATC) code and filter drug records with at least one drug feature. Finally, five types of triples are constructed: <drug, hasTransporter, Transporter>, <drug, hasADR, ADR>, <drug, hasEmzyme, Emzyme>, <drug, hasTarget, Target>, <drug, hasIndication, Indication>; The tuples are stored in the Neo4j graph database to obtain a visual knowledge graph, as shown in Figure 2. The map contains a total of 8001 drugs, 5453 ADRs and 158133 triples, as shown in Table 2.

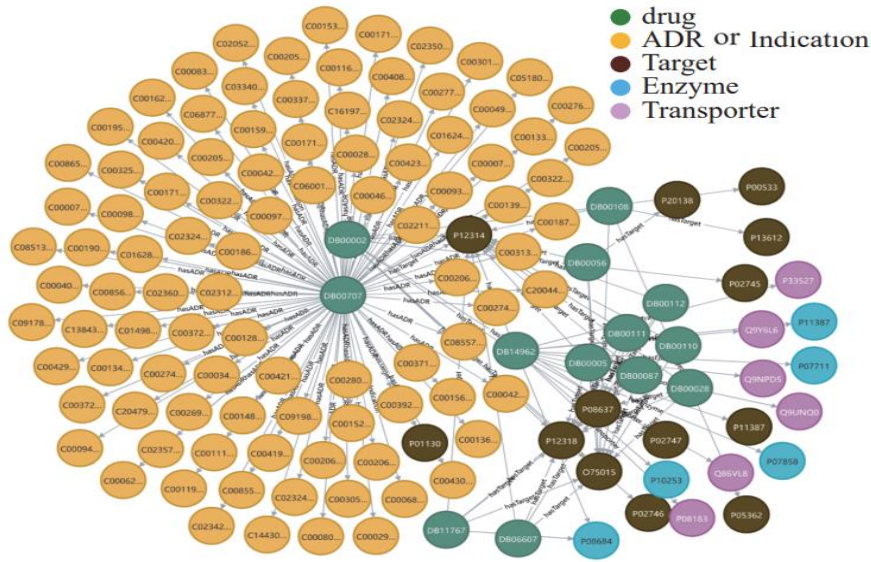

Figure 2. Local entities and relationships in the knowledge graph

Table 2 Entities, Relationships and Quantities Included in the ADR Knowledge Graph

| $h$ | $h$ number/type | $r$ | $t$ | number of $t$ | triplet type | Number of triples |
|---|---|---|---|---|---|---|
| drug |  | hasTarget | Target | 4709 | < drug, hasTarget, Target > | 18172 |
| drug |  | hasTransporter | Transporter | 268 | < drug, hasTransporter, Transporter > | 3061 |
| drug | 8001 | hasEnzyme | Enzyme | 433 | < drug, hasEnzyme, Enzyme > | 5149 |
| drug |  | hasIndication | Indication | 2642 | < drug, hasIndication, Indication > | 12491 |
| drug |  | hasADR | ADR | 5453 | < drug, hasADR, ADR > | 119260 |

| Total | —— | —— | —— | —— | —— | 158133 |

## B. Knowledge graph embedding model

Knowledge graph embedding technology is gradually used in prediction research[18], among which the DistMult[19] model and HolE[20] model based on tensor decomposition are the most widely used. The DistMult model describes the semantic correlation between entities through bilinear transformation between entities, where the head entity and tail entity are represented by vectors $h$ and $t$ respectively, and the relationship is represented by vector $r$; the relationship matrix $M_r = diag(r)$ pair Pairwise interactions between latent factors are modeled using $f_r(h,t) = h^T M_r t$ as the scoring function. The HolE model is based on the DistMult model and introduces circular correlation operations between entities to capture the compositional representation of pairs of entities. It uses $f_r(h,t) = r^T(h * t)$ as the scoring function, where * is the circular correlation. Operation. Both of above two embedding models take minimizing the scoring function as the goal to obtain effective embedding vectors of entities and relationships.

## C. CNN classification model

A CNN model with 2 convolutional layers and 4 fully connected layers is researched and designed, as shown in Figure 3. Since the calculation efficiency and convergence speed of the ReLU activation function are much higher than those of sigmoid, Tanh and other functions; therefore, both the convolutional layer and the fully connected layer use the ReLU activation function. At the same time, in order to maintain the same distribution of inputs of each layer of neural network and improve network optimization efficiency, the convolutional layers all use batch normalization. The specific parameters of the model are shown in Table 3. This paper uses the binary cross entropy shown in Equation (1) as the loss function for model training, where $n$ is the total number of training samples, $y_i$ is the true label of sample $i$, and $y_i$ is the probability that sample $i$ is predicted to be category "1" value; through model training, obtain the optimal values of parameters W and b.

$$J(W,b) = -\frac{1}{n}\sum_{i}^{n}[y_i log \hat{y}_i + (1 - y_i)\log(1 - \hat{y}_i)] \quad (1)$$

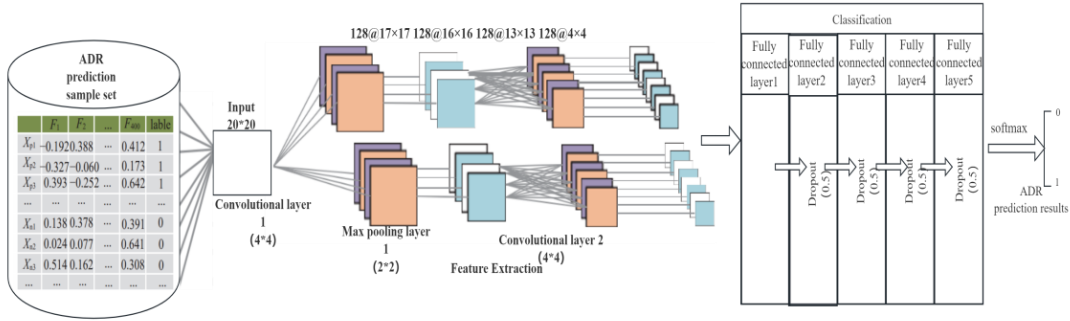

Figure 3. CNN model structure diagram for ADR prediction

Table 3. Parameters of CNN model

| Input | Layers | Cores | Number of cores | Step size | Output |
| --- | --- | --- | --- | --- | --- |
| 20×20 | Convolution layer 1 | 4×4 | 128 | 1 | 17×17×128 |
| 17×17×128 | max pooling layer 1 | 2×2 | —— | 1 | 16×16×128 |
| 16×16×128 | Convolutional layer 2 | 4×4 | 128 | 1 | 13×13×128 |
| 13×13×128 | max pooling layer 2 | 4×4 | —— | 3 | 4×4×128 |
| 4×4×128 | Fully connected layer 1 | —— | —— | —— | 1 024 |
| 1024 | Fully connected layer 2 | —— | —— | —— | 256 |
| 256 | Fully connected layer 3 | —— | —— | —— | 64 |
| 64 | Fully connected layer 4 | —— | —— | —— | 32 |
| 32 | Fully connected layer 5 | —— | —— | —— | 2 |

Using Logistic Regression (LR), K-Nearest Neighbor (KNN), Decision Tree (DT), Random Forest (RF), Naive Bayes (NB), Comparative analysis of six benchmark models including Gradient Boosting Decision Tree (GBDT), which are widely used in ADR prediction[21].

## IV. EXPERIMENT AND RESULT ANALYSIS

### A. Model Evaluation Indicators

This project uses the confusion matrix to calculate the recall rate (Recall), accuracy rate (ACC), precision rate (Precision, P), $F_1$ value ($F_1$- Score, F1) and area under the curve (AUC) as a model evaluation indicators.

### B. Knowledge graph embedding and sample vector representation

The embedding operation is based on the Python language and is implemented by calling the AmpliGraph tool library. Before the embedding operation, it is necessary to determine the training set and test set of the ADR prediction model; the training set is used for the knowledge graph embedding operation and ADR prediction model training, and the test set is used to evaluate the prediction performance of the ADR prediction model. There are 119,131 positive samples and 12,435 negative samples in the knowledge graph (see Table 4). Since the number of positive and negative samples differs by one order of magnitude, based on the total number of negative samples, the negative samples are randomly divided into 11,305 training samples and 1,130 test samples at a ratio of 9:1, and randomly selected from the positive samples. Take 1,130 as test samples; then the test set contains 1,130 positive and negative samples; the training set includes 118,001 positive samples and 11,305 negative samples. In order to solve the imbalance problem of training set samples, oversampling is used to copy the negative samples 10 times. The sample division results are shown in Table 4.

Table 4. Data Used For Kg Embedding and ADR Classifier Training and Testing

| Triple type | Knowledge graph embedding/piece | Classifier training/piece | Classifier test/piece | Total/piece |
|---|---|---|---|---|
| < drug, hasTarget, Target > | 18162 | 0 | 0 | 18162 |
| < drug, hasTransporter, Transporter > | 3102 | 0 | 0 | 3102 |
| < drug, hasEnzyme, Enzyme > | 5080 | 0 | 0 | 5080 |
| < drug, hasIndication, Indication > | 11305 | 11305*10 | 1130 | 12435 |
| < drug, hasADR, ADR > | 118001 | 118001 | 1130 | 119131 |

In the process of knowledge graph embedding, this paper uses different embedding strategies to obtain embedding vectors. And use $h_D, t_A, t_I$ to represent the embedding vectors of the entities drug, ADR and Indication respectively, and subtract the tail entity vector from the head entity vector to construct the representation vector of the positive and negative samples of the ADR classifier, as shown in Table 5. $X_p$ and $X_n$ are used to represent positive samples and negative samples respectively, where $X_p$ corresponds to the "Drug-ADR" combination and $X_n$ corresponds to the "Drug-Indication" combination. $X_n$ and $X_p$ together constitute the experimental data set of the classifier.

Table 5. Representation Vector of Partial Samples of ADR Classifier (Distmult, Dim=20)

|   | Drug | ADR | 1 | 2 | 3 | … | 18 | 19 | 20 | Label |
|---|---|---|---|---|---|---|---|---|---|---|
| 0 | DB00513 | C0023890 | -0.2064 | 0.4520 | -0.3854 | … | 0.1960 | 0.3563 | 0.2631 | 0 |
| 1 | DB01320 | C0011206 | -0.2065 | 0.4520 | -0.3866 | … | 0.1958 | 0.3571 | 0.2639 | 0 |
| 2 | DB01241 | C0030305 | -0.2072 | 0.4521 | -0.3865 | … | 0.1952 | 0.3580 | 0.2633 | 0 |
| 3 | DB01141 | C0239295 | -0.2091 | 0.4528 | -0.3874 | … | 0.1962 | 0.3588 | 0.2658 | 0 |
| 4 | DB01059 | C0033581 | -0.2063 | 0.4500 | -0.3861 | … | 0.1958 | 0.3571 | 0.2629 | 0 |

### C. Comparative Analysis of Embedded Dimensions

This paper explores the impact of different embedding strategies on the prediction performance of the benchmark ADR classification model on the test set by combining different embedding models and different embedding dimensions (10 to 800). As shown in Figure 4, under different embedding models, as the embedding dimension increases, the AUC values of each benchmark model on the test set gradually increase; and the ACC and F1 index values also fluctuate to varying degrees; Recall The value does not increase significantly and is relatively stable. However, when the embedding dimension is greater than 400, the AUC, ACC, and F1 index values of each benchmark model tend to be stable. Through comprehensive analysis, appropriately increasing the embedding dimension can improve the prediction performance of the ADR classification model to a certain extent. At the same time, in order to avoid overfitting of the classifier and waste of experimental hardware equipment resources, this paper selects 400 dimensions as the optimal embedding dimension and combines the CNN model for ADR prediction.

### D. Comparative Analysis of Classification Models

Based on the Python language, scikit-learn and the deep learning framework Tensorflow2.0 are used to develop the ADR classification model. The six benchmark models will use default parameters. The fixed embedding dimension is 400 dimensions. The representation vector of the sample is obtained through the embedding model and input into the ADR classification model for training and prediction. The prediction results of each classification model on the test set are shown in Table 6. Comprehensive analysis found that under the DistMult embedding model, the AUC value of the CNN classification model on the test set was 0.942, which was better than all benchmark models.

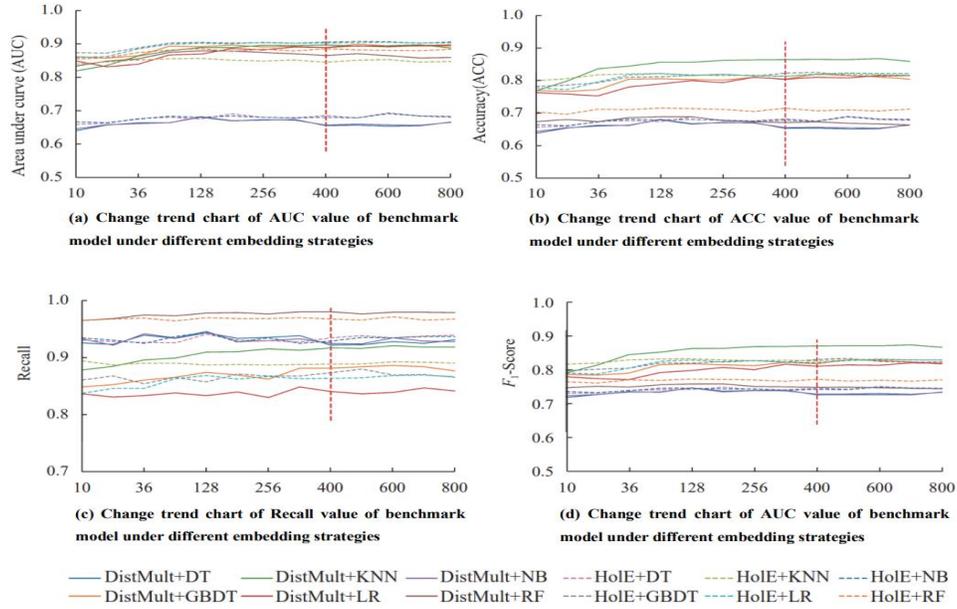

Figure 4. The performance of each baseline ADR classification model on the test set with different embedding dimension

Table 6. Comparison of ADR Prediction Models When the Embedding Dimension Is 400

| Embedded model | Classifier | AUC | ACC | P | F1 | Recall |
|---|---|---|---|---|---|---|
| DistMult | LR | 0.889 | 0.804 | 0.784 | 0.811 | 0.841 |
| | RF | 0.866 | 0.672 | 0.607 | 0.749 | 0.98 |
| | KNN | 0.897 | 0.864 | 0.83 | 0.871 | 0.917 |
| | GBDT | 0.889 | 0.805 | 0.765 | 0.819 | 0.882 |
| | DT | 0.656 | 0.655 | 0.601 | 0.727 | 0.922 |
| | NB | 0.658 | 0.657 | 0.602 | 0.729 | 0.925 |
| | CNN | 0.942 | 0.847 | 0.794 | 0.86 | 0.938 |
| HolE | GBDT | 0.906 | 0.823 | 0.793 | 0.832 | 0.873 |
| | DT | 0.687 | 0.684 | 0.623 | 0.747 | 0.934 |
| | NB | 0.681 | 0.679 | 0.62 | 0.743 | 0.929 |
| | CNN | 0.927 | 0.843 | 0.802 | 0.853 | 0.91 |

E. Model Stability Assessment

The study used 5 repeated experiments to evaluate the stability of the CNN model. Specific steps: (1) Set random seeds and construct training sets and test sets; (2) Use the "DistMult model + 400 dimensions" combination strategy for embedding operations; (3) Use the obtained sample representation vectors for CNN classification model training and predict. The results are shown in Table 7. The average AUC of the CNN model in this article is 0.957; the average $F_1$ value is 0.891, and the average Recall value is 0.914, each indicator value fluctuates less. At the same time, the ROC curve (shown in Figure 5) is also stable, indicating that the CNN model developed in this article has high stability.

Table 7. The Performance of the CNN Model on the Test Set For Five Repeated Experiments

| Experiments | Random | Evaluation indicators | | | | |
|---|---|---|---|---|---|---|
| | | AUC | ACC | P | F1 | Recall |
| Experiment 1 | 18 | 0.957 | 0.884 | 0.86 | 0.888 | 0.917 |
| Experiment 2 | 24 | 0.965 | 0.872 | 0.843 | 0.886 | 0.926 |
| Experiment 3 | 36 | 0.952 | 0.887 | 0.875 | 0.889 | 0.903 |
| Experiment 4 | 40 | 0.962 | 0.897 | 0.887 | 0.898 | 0.909 |
| Experiment 5 | 48 | 0.959 | 0.89 | 0.871 | 0.892 | 0.914 |
| Average value | — | 0.959 | 0.886 | 0.867 | 0.891 | 0.914 |

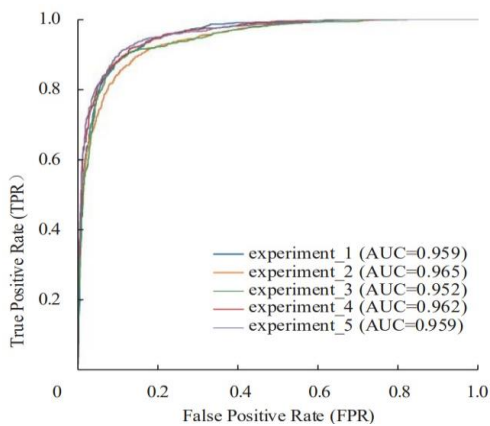

Figure 5. ROC curve of five repeated experiments of CNN model

## F. Prediction Model Verification

This project uses real-world data to test the effectiveness of the CNN model. Using "kidney injury" or "kidney injury" as keywords, we randomly searched for relevant ADR studies in literature databases such as PubMed, and obtained 3 "Drug-ADR" combinations that were not included in the SIDER database; these were used as Input, use CNN model to make predictions. The results show (shown in Table 8) that the average probability of a real sample being predicted as "positive" is 0.967, indicating that the CNN model in this article can effectively predict samples outside the experimental sample set.

Table 8. Prediction Results of Drug-ADR Pairs In Literature Through CNN Model

| Drug | ADR | Predicted probability | Documentary evidence |
| --- | --- | --- | --- |
| tazobactam＜DB01606＞ | Acute kidney injury(C2609414) | 0.921 | Literature[22] |
| Osimertinib＜DB09330＞ | Acute kidney injury(C2609414) | 0.982 | Literature[23] |
| Nivolumab＜DB09035＞ | Kidney disease(C0022658) | 0.999 | Literature[24] |

1. ＜ ＞ is the ID of the drug in the DrugBank database;
2. ( ) is the MedDRA code corresponding to the ADR.

## G. Comparative Analysis With Advanced Research

Due to the current lack of standard data sets for testing the performance of ADR prediction models, this article will compare it with related typical studies in terms of the drugs covered, the number of ADR types, and the AUC value of the prediction model (see Table 9). Through comparative analysis, the AUC of the CNN model developed in this article is higher than the results provided by related research, and the prediction performance is better. At the same time, the experimental data set of this article contains 7920 drugs and 5454 ADRs, covering more drug information than most similar studies. In addition, most previous studies need to build a separate prediction model for each ADR, which increases the workload of the ADR prediction task. In contrast, this paper builds a drug knowledge graph and uses knowledge graph embedding technology to encode entities such as drugs and ADRs into Feature vector; Finally, a unified CNN model is used to predict each " " combination to evaluate the probability that the combination has the "hasADR" relationship, which greatly reduces the number of models. The study by Zhang et al.[12] used a similar method for ADR prediction, but it covered only 3,632 drugs and showed a relatively low AUC value; the study by Joshi et al. [11] was reported in the literature[12] added drug pathway (Pathways) and gene (Gene) features, but the average AUC of its ADR prediction model is only 0.912, and there is still room for improvement. This paper develops a higher-performance ADR prediction model by selecting more representative drug features.

Table 9. Comparison With Advanced ADR Prediction Models

| Research | Number/type of drugs | Feature category | Number of ADR | Label data source | Whether to introduce knowledge graph | Classification model | Average AUC |
| --- | --- | --- | --- | --- | --- | --- | --- |
| This Project | 7 920 | Target, Indication, Transporter, Enzyme, ADR | 5 454 | SIDER | Yes | CNN | 0.955 |
| Literature[11] | 7 219 | ADR, Indication, Target, Pathway, Gene | 5 469 | SIDER | Yes | DNN | 0.912 |
| Literature[12] | 3 632 | Target, Indication, ADR | 5 589 | SIDER | Yes | LR | 0.86 |
| Literature[14] | 1 430 | Chemical structure, Side effect | 1 766 | SIDER | No | CNN | 0.919 |

## V. CONCLUSION

In view of the problems of previous ADR prediction model research, such as low prediction accuracy and heavy workload due to the need to build separate classifiers for each ADR, this project simplifies the prediction of different types of ADR into a binary classification problem and develops a model based on knowledge graph embedding and depth. Learned CNN prediction model. By comparison, the prediction model in this article is more accurate than the predictions of existing studies. In addition, the validity and feasibility of the model prediction results are verified through real-world data, and it is expected to play a core auxiliary role in clinical safe medication in the future. The next step of research will consider using similar methods to study the potential adverse reactions of different drug interactions; or be patient-centered, assess potential risk factors that lead to ADRs in clinical patients, and predict patients' specific drug use situations. In the future, the relationship between drug morphology and potential side effects can be further studied based on existing reinforcement learning algorithms[25] and image segmentation[29].